# Accurate and Efficient Measurements of IP Level Performance to Drive Interface Selection in Heterogeneous Wireless Networks

S. Salsano, F. Patriarca, F. Lo Presti, P. L. Ventre, V. Gentile

**Abstract**— Optimal interface selection is a key mobility management issue in heterogeneous wireless networks. Measuring the physical or link level performance on a given wireless access networks does not provide a reliable indication of the IP connectivity, delay and loss on the (bidirectional) paths from the Mobile Host to the node that is handling the mobility, over different heterogeneous networks. In this paper, we propose, implement and analyze mechanisms for connectivity check and performance (network delay and packet loss) monitoring over IP access networks. We evaluate the accuracy and timeliness of the performance estimates and provide guidelines for tuning up the parameters. From the implementation perspective, we show that using application level measurements is highly CPU intensive, while a kernel based implementation has comparably a very low CPU usage. The Linux kernel implementation results in an efficient use of batteries in Mobile Hosts and intermediate Mobility Management Nodes can scale up to monitoring thousands of flows. The proposed solutions have been implemented in the context of a specific mobility management solution, but the results are of general applicability. The Linux implementation is available as Open Source.

**Index Terms**—Network monitoring, wireless networking, mobility management, vertical handover.

✦

## 1 INTRODUCTION

Several solutions have been proposed in the last years for mobility management in IP based heterogeneous networks, working at different protocol levels, from layer 2 up to application level [1], [2], [3]. Nevertheless, mobility management is still an open issue for research and standardization.

We consider a scenario in which the terminals have multiple and heterogeneous wireless interfaces (e.g. WiFi, 3G/4G, WiMax) that can be active at the same time. Terminals can conveniently switch from one interface to another (handover) to optimize some suitable network performance parameters, e.g., round trip time and packet loss ratio, with the goal to improve the application level performance and the user experience in general.

In such a scenario, the handover decision process, that is the determination of when and on which interface to switch, plays a key role. Several solutions for the handover decision process have been proposed and evaluated in the literature (see [4] for a comprehensive review and discussions of the research issues). The Mobile Host and/or the network can take into account several factors to drive the handover process, from the received signal strength on the radio interface, to the cost of connectivity, the desired QoS, the battery usage and so on. Among these, it is relatively easy to evaluate the radio link performance on a given wireless access network. Unfortunately, the radio link performance provides neither a reliable indication on availability of the end-to-end connectivity nor a meaningful information on the level of service provided to the Mobile Host and its applications, should the network flows be handed over that wireless access network. For example, consider a Mobile Host connected to a WiFi hot-spot and to a 3G network. The link level performance on the wireless link between the Host and the WiFi access point can be very high, while at the same time the connectivity of the WiFi access point to the Internet can be very poor (or even not present). Clearly, in this scenario it would not be wise to take the handover decision between the WiFi and the 3G access network only relying on the link level measurements since the terminal will experience poor application level performance, despite the good WiFi link level quality. To overcome these limitations, we consider the adoption of *connectivity checks* and *performance measurements* at the IP level on the path from the Mobile Host up to the intermediate *Mobility Management Node* that handles the mobility (or up to the Correspondent Host if the mobility is handled end-to-end).

Performing a continuous connectivity check and gathering the IP performance measurements in a timely, effective and efficient way is a demanding task. The impact on these procedures on the processing load of Mobile Hosts and of Mobility Management Nodes and on the network load needs to be carefully assessed. From the analysis of the literature on mobility management (see the surveys [1], [2], [3]) we believe that these aspects have not been adequately covered so far. Most of the papers deal with architectural and protocol aspects of handover management, and/or focus on the performance of the handover procedure itself, but do not address the measurements procedures themselves which are needed to drive the interface selection and their computational and network load impact.

There has been a considerable amount of work on IP level performance measurements and several tools are available (see [5]) to estimate network delays, packet loss ratios, and

---

- S. Salsano, F. Patriarca, P.L. Ventre, V. Gentile are with the Electronic Engineering Dept., University or Rome Tor Vergata, E-mail: salsano@ieee.org, fabio.patriarca.2@uniroma2.it.
- F. Lo Presti is with the DICII Department, University or Rome Tor Vergata, E-mail: lopresti@disp.uniroma2.it.

available bandwidth. This work has been done mostly from the perspective of network management. Based on our analysis of requirements for the needed connectivity check and performance measurements procedures (see section 3), we realized that no existing tools provides a solution which is efficient for our purposes and can be easily integrated into a handover management system for Mobile Host and Mobility Management nodes. For this reason, we designed and implemented the solutions proposed in this paper. Specifically, the main contributions of our work are:

- Design of optimized connectivity check and IP network performance measurements procedures for Round Trip Time and packet loss ratio.
- A theoretical analysis of tradeoffs for the connectivity check procedure between responsiveness and processing/network load, with the identification of optimal parameter selection.
- An Open Source implementation of the proposed procedures, with different approaches (user space / kernel space) [6][7]. The code can be reused and integrated into any mobility management solution.
- Evaluation of the processing load of the different solutions with real measurements taken in a test bed
- Verification of the accuracy and timeliness of the proposed network performance measurements, based on the real implementation.

We base the implementation and analysis of the proposed mechanisms on a specific mobility management solution called UPMT (Universal Per-application Mobility Management using Tunnels) [8]. Nevertheless, *our findings are of general value* and not restricted to the UPMT solution. The proposed mechanism and results are relevant to all mobility management solutions that combine heterogeneous networks using IP (e.g. Mobile IP [9], HIP – Host Identity Protocol [10], DMM – Distributed Mobility Management [11][12]). In facts, all these solutions share the need of performing connectivity checks and network performance monitoring.

The paper is organized as follows. Section 2 introduces the UPMT mobility management solution and its usage scenarios. Section 3 analyses the requirement for the connectivity check and network performance monitoring procedures (dealing with packet delay and loss). Section 4 describes the design of the proposed procedures. Sections 5 deals with the implementation aspects, considers some optimizations and provides an evaluation of the processing cost for different implementation choices. In section 6, the accuracy and timeliness of the mechanisms are discussed. Section 7 reports an analysis of related work and finally conclusions are drawn in section 8.

## 2 UPMT BASICS AND USAGE SCENARIOS

UPMT is a solution for mobility management over heterogeneous networks based on IP in UDP tunneling. In this section we shortly recall its main features, further details can be found in [6][8][13]. A Mobile Host establishes IP in UDP tunnels over its active network interfaces with its "correspondent" UPMT node. This correspondent UPMT node can be an "Anchor node" (see Fig. 1) or a correspondent UPMT aware Host (see Fig. 2). The UPMT solution can be applied to different scenarios, we consider two of them in this paper. The first scenario, called *Internet access* is shown in Fig. 1. A Mobile Host is connected to a mobility management node denoted as *Anchor Node* via different access networks and it has to choose the "best" access network over time. The second scenario is called *peer-to-peer multi-access*. It assumes that a set of devices with multiple network interfaces can communicate in a peer-to-peer fashion and want to select the best network interfaces to be used dynamically. A particular example of this scenario is a mobile ad-hoc network in which the nodes have multiple WiFi interfaces, as shown in Fig. 2.

The tunnels are used to exchange the IP packets according to the format shown in Fig. 3. The "external" packet has IP source and destination addresses corresponding to the IP addresses of the interfaces of the Mobile Host and of the correspondent UPMT node. The internal encapsulated packet can keep the same IP source and destination addresses irrespective of the interfaces used for sending and receiving the packet. This allows seamless handovers of flows among multiple tunnels setup between the Mobile Host and the correspondent UPMT node.

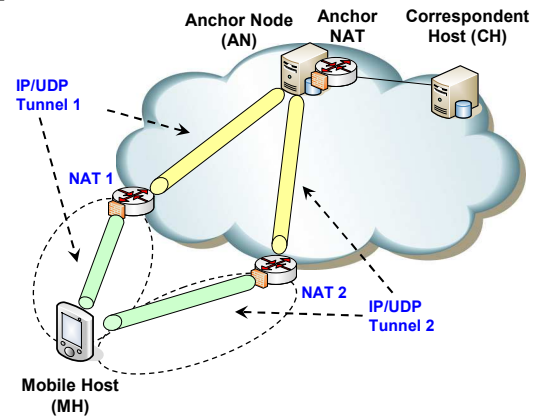

Fig. 1 Internet Access scenario

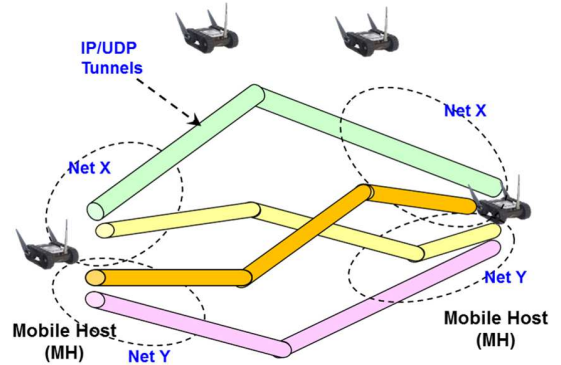

Fig. 2 Peer-to-peer multi-access scenario

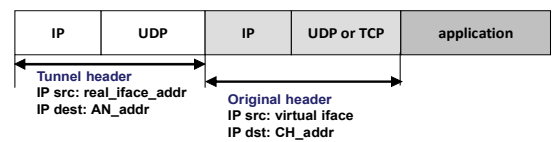

Fig. 3 UPMT packet format

In our Linux implementation of UPMT, the UPMT kernel module provides a virtual interface called UPMT0 as a regular networking device, as shown in Fig. 4. A "virtual" IP address can be assigned to it and the legacy applications will see a standard networking device. The UPMT encapsulation and mobility management is completely transparent for the applications that can use plain sockets to communicate.

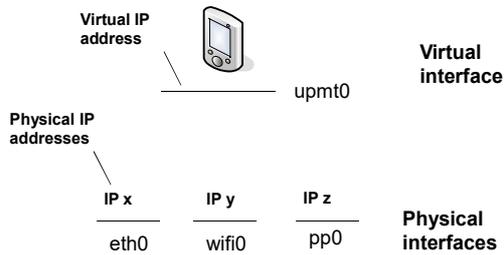

Fig. 4 UPMT virtual interface vs. physical interfaces

Considering for example the Internet access scenario, if a tunnel over a given access network is used and the connectivity towards the Anchor Node through such tunnel fails, the active flows should be immediately handed over another tunnel on the second access network. If the failure happens on the radio access interface, it could be detected by monitoring of the radio link. If the failure happens on any node or link behind the radio access point in the path toward the Anchor Node, it is undetectable using the radio link monitoring. The same applies to the peer-to-peer multi access scenario when considering the end-to-end tunnels among the mobile hosts: the radio link monitoring is not enough to assess the liveliness and the quality of the end-to-end connection.

Therefore, the only option is to perform a continuous monitoring at IP level, checking which tunnels provide connectivity towards the correspondent UPMT nodes and what is the performance (delay and loss ratio) of the connected tunnels. Efficient mechanisms are needed to detect a sudden loss of connectivity or a sharp decrease in performance on a connected tunnel.

## 3 ANALYSIS OF THE REQUIREMENTS

Let us proceed from the ground-up by analyzing the requirements for the connectivity check and performance monitoring procedures. For simplicity, we will also globally refer to these procedures as *Keep alive* procedures (because they can rely on sending periodic messages referred to as keep alive messages). The Keep alive procedures are required to perform at the same time: i) the evaluation of Round Trip Time (RTT [ms]); ii) the evaluation of One Way Loss (OWL) ratio in the two directions; iii) the connectivity check, used to monitor the state of the connectivity on the paths over the different interfaces and to detect failure conditions as soon as possible. The RTT is a "bidirectional" delay measurement, as it takes into account the transit delay in the tunnel in both directions. For most services, like conversational real-time communications, client-server requests, TCP based data transfer, the RTT is the most important performance parameter (as shown in [14], TCP throughput is proportional to $1/(RTT\sqrt{p\_loss})$). Only for a small subset of services like unidirectional real-time broadcast, it could be rather of interest to measure the One Way Delay (OWD) in one of the directions. Unfortunately, this would require clock synchronization between the two ends of the tunnel. Therefore, in this work we only consider RTT measurements.

The key specific functional and non-functional requirements that we have envisaged are listed in Table 1. We highlight the importance, among the non-functional requirements, of minimizing the CPU load and the amount of state information. In Mobile Hosts this corresponds to minimizing the power consumption, while in Mobility Management nodes this will maximize the number of Mobile Hosts supported with a given amount of resources.

Table 1 Functional and Non-Functional requirements

| Functional requirements | |
|---|---|
| 1 | Detection of loss of connectivity (the responsiveness should be configurable) |
| 2 | Measurement of delay (round trip) |
| 3 | Measurement of one way loss in the two directions |
| 4 | Configurable time accuracy in the evaluation of the performance parameters |
| 5 | Estimation of averages of performance over configurable time scales |

| Non-Functional requirements | |
|---|---|
| 1 | Resilience to packet loss and jitter for all measurement procedures |
| 2 | Minimal memory footprint (i.e. minimize state information for each interface/remote end-point to be monitored) |
| 3 | Minimal CPU load |
| 4 | Combined solution for all the measurements to minimize the packets to be sent and processed. |
| 5 | Self-contained code, no dependency on external libraries or modules |

In general, the connectivity check and performance monitoring can be done using an *active* approach (i.e. sending probe packets) or with a *passive* approach (i.e. trying to infer connectivity status and tunnel performance from the observation of existing traffic). In theory, the passive approach is preferable because it does not introduce additional traffic into the network. On the other hand, inferring the performance from existing traffic can be more complex and CPU intensive and in any case it is not feasible to rely on purely passive measurements for certain tasks like connectivity check. Considering the fundamental requirement that measurements and connectivity check must be available also in absence of traffic, we decided to focus on the *active* approach.

The measurements collected by the Keep Alive procedures will be processed by the entities that performs the handover decision in a given mobility management architecture. The decision process is logically separated and independent from the Keep alive procedures; the premise is that the IP level measurements of the paths over the different wireless access network provide valuable information to take the optimal handover decisions for all the types of applications (and also different decisions for each type of application if the mobility management architecture supports this approach). The proposed approach gives the possibility to take handover decisions based on the combination of RTT and loss metrics.

## 4 DESIGN OF CONNECTIVITY CHECK AND PERFORMANCE MONITORING PROCEDURES

In this section we describe the proposed procedures for the evaluation of RTT, OWL and for performing the connectivity check. The three procedures are combined in a common framework, in which for each tunnel[1], one end of the tunnel plays the *client* role while the other end plays the *server* role. The client role is taken by the end that starts the tunnel establishment with a tunnel setup request. The other end, that receives the tunnel setup request message, will play the server role. The client-end periodically executes the keep alive procedure each $T_{KA}$ seconds by sending a *probe request* packet towards the server-end for each active tunnel. The server-end sends back a *probe response* packet. The procedures described in this section can be applied to both scenarios described in section 2. The two involved entities (client end and server end) will be respectively a Mobile Host and a Mobility Management node in the Internet Access scenario, or two peers in case of an end-to-end Mobility Management solution.

### 4.1 RTT evaluation

We assume that both ends are interested to evaluate the RTT. With reference to Fig. 5, we define as $t_{Sc}$ the time instant when the probe request is sent by the client-end, $t_{Rs}$ the time instant when the server-end receives the probe request, $t_{Ss}$ the time instant when the server-end sends the probe response, $t_{Rc}$ the time instant when the client receives the probe response. Note that the client and server clocks do not need to be synchronized, therefore $t_{Sc}$ and $t_{Rc}$ represent the times as measured by the client clock, while $t_{Rs}$ and $t_{Ss}$ the times as measured by the server clock.

The probe request messages include 3 parameters:

$t_{Sc}(k)$, $t_{Ss}(k_{prev})$, $\Delta t_C(k) = t_{Sc}(k) - t_{Rc}(k_{prev})$

where $t_{Ss}(k_{prev})$ and $t_{Rc}(k_{prev})$ represent the most recently received values for these state variable.

The probe response messages include 3 parameters:

$t_{Ss}(k)$, $t_{Sc}(k)$, $\Delta t_S(k) = t_{Ss}(k) - t_{Rs}(k)$

In this way, both ends of the tunnel can evaluate the RTT delay from the probe packets without keeping a state information, as follows:

On the client-end:
$RTTc(k) = t_{Rc}(k) - t_{Sc}(k) - \Delta t_S(k)$
On the server-end:
$RTTs(k) = t_{Rs}(k) - t_{Ss}(k_{prev}) - \Delta t_C(k)$

The client-end needs to explicitly store the $t_{Ss}$ state variable until it sends the next probe request, which typically happens on a timer basis. The server-end does not need to store the $t_{Sr}$ state variable because the probe response is sent immediately after receiving the probe request and this state variable is local to the procedure that handles the probe request. In section 5.3 we will propose an optimization of the procedure, in which the probe response can be sent after a delay, in such case the explicit storage of the $t_{Sr}$ state variable for later retrieval is needed.

RTT(k) can potentially assume a different value each time a new probe packet is received. This information can be accumulated using an EWMA (Exponentially Weighted Moving Average) procedure so that a single state variable per tunnel can represent the RTT performance of the tunnel during the recent past (e.g. the last minute or so). The proposed EWMA algorithm is explained in Appendix I. It takes into account that the samples to be averaged are available at time intervals that are not regular, due to the variation of the RTT itself and that some RTT samples could be missing (because of the loss of probe packets). The algorithm is characterized by its *time constant* $\tau_{RTT}$. A smaller time constant means that the EWMA reacts faster to the changes of the estimated parameter, but also that it takes into account only the more recent values of the parameter.

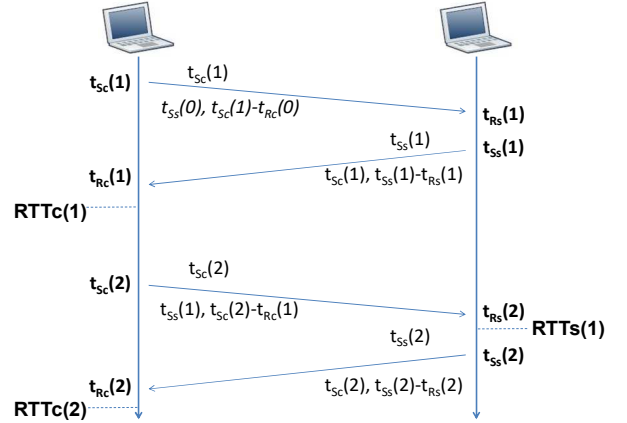

Fig. 5 Time sequence for RTT evaluation procedure

Actually, it is also possible to maintain different EWMA state variables with different time constants in order to accumulate the information at different time scales (for example a shorter time scale in the order of few seconds and a relatively longer time scale in the order of few tens of seconds or few minutes). The RTT state variable(s) are needed in both sides if both sides are interested in evaluating the RTT.

In the described solution, both the client and the server-ends independently evaluate the RTT. There can be scenarios in which only the client-end is interested to evaluate the RTT, for example in a client-server application driven by the client. In this case, no state information is needed in the server-end, minimizing the server side resources needed to handle the Keep-Alive procedure.

### 4.2 Loss evaluation (One Way Loss)

Let us consider now the estimation of loss ratio. We define the One Way Loss (OWL) ratio as the fraction of lost packets with respect to the transmitted packet. This can be measured both for the packets that are transmitted from the client-end to the server-end of a tunnel and from the server-end to the client-end. The former will be denoted as $OWL_c$, the latter as $OWL_s$. The time interval $T_L$ [s] over which this percentage is evaluated is arbitrary and characterizes the OWL measurements. The exchange of probe requests/responses happens on a periodic basis with period $T_{KA}$ [ms]. The OWL evaluation interval $T_L$ is chosen as a multiple of $T_{KA}$: $T_L = N*T_{KA}$. The factor N should be chosen so that the number of transmitted packets during the interval allows evaluating a meaningful ratio. The OWL can only be evaluated when receiving the first probe response (for the client-end), or the first probe request (for the server-end) after the $T_L$ expiration. The sequence of evaluated OWL values will be denoted as OWL(m).

We assume that both ends are interested to evaluate the OWL. With reference to Fig. 6, we define as Sc and Rc the total number of packets sent and received by the client-end on the tunnel, Ss and Rs the number of packets sent and received by the server-end. These counters include both the data and the probe packets. More precisely, the client-end increases the Sc variable for each packet sent in the tunnel and the Rc variable for each received packet. Likewise, the server-end increases the Ss variable for each sent packet and the Rs variable for each received packet.

---

[1] In the terminology we refer to a tunnel based mobility solution like UPMT, but the concepts can be adapted to solutions not using tunnels.

The probe request messages include 3 parameters:
  $Sc(k)$, $Ss(k_{prev})$, $Rc(k_{prev})$
The probe response messages include:
  $Ss(k)$, $Sc(k)$, $Rs(k)$

After the $T_L$ timer expires, both ends of the tunnel evaluate the OWL ratio, as soon as they receive a probe packet. The received probe packet has the index k, and will produce the evaluation of the $m^{th}$ OWL value.

On the client-end:
  $OWLc(m) = 1 – ((Rs(k) – Rs\_last) / (Sc(k) – Sc\_last))$
  $Sc\_last \leftarrow Sc(k)$
  $Rs\_last \leftarrow Rs(k)$

On the server-end:
  $OWLs(m) = 1 - ((Rc(k) – Rc\_last) / (Ss(k) – Ss\_last))$
  $Ss\_last \leftarrow Ss(k)$
  $Rc\_last \leftarrow Rc(k)$

Where Sc_last and Rs_last on the client-end, and Ss_last and Rc_last on the server-end respectively store the Sc, Rs, Ss and Rc values as they will be needed for the next evaluation of OWL. Obviously all _last variables are initialized to zero for the first OWL evaluation.

Considering that probe packets can suffer a variable delay or can be lost, the OWL evaluation will not happen exactly every $T_L$. It can even occur that no probe packets are received for a whole $T_L$ duration, in this case the OWL evaluation for the given interval will be missing, but this is not critical as the OWL evaluation in the next $T_L$ will take into account the packets that have been lost. In general, the sequence number m of the evaluated OWL values will be such that m <= k/N (where $T_L$ = N*$T_{KA}$ and the equality holds if all probes have been received).

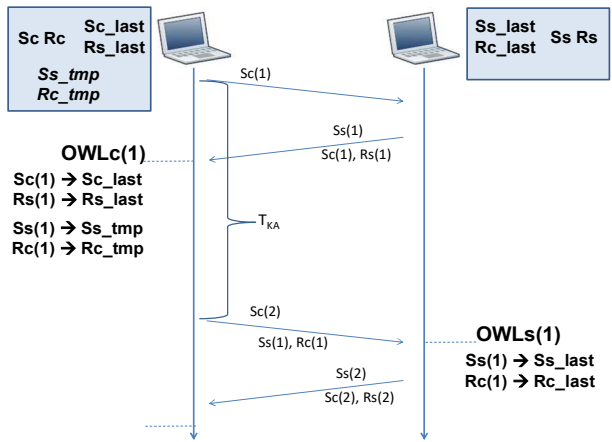

Fig. 6 Time sequence for OWL evaluation procedure

Both ends needs to maintain 4 state variables: 2 variables to continuously update the number of sent and received packets and 2 variables which will be updated every time the OWL is evaluated and are denoted with the _last pedix. In addition, the client-end needs to explicitly store the Ss and the Rc state variables, respectively in the Ss_tmp and Rc_tmp variables, until it sends the next probe request, which happens on a timer basis.

It is important to observe that the client-end and server-end clocks do not need to be synchronized because each end is able to evaluate independently the OWL over each $T_L$ period. In principle, the time period $T_L$ for the OWL evaluation can even be different in the client-end ($T_{Lc}$) and in the server-end ($T_{Ls}$), but for simplicity we considered a single period ($T_L=T_{Lc}=T_{Ls}$).

Rather than keeping the whole sequence of OWL(m), it is possible to accumulate them using an EWMA procedure, exactly as already discussed for the RTT measurement. In this way, a single state variable per tunnel represents the OWL performance at a given time scale (and it is possible to keep multiple state variable using different time constants in order to consider different time scales).

Finally, we observe that a *Round Trip Loss* (RTL) metric could be used instead of OWLc and OWLs, if it is acceptable to have an aggregated estimation of the loss in the two directions. As discussed in Appendix III, this brings a simplification of the measurement procedure and a reduction of the required state information.

### 4.3 Connectivity check

In the *active* procedure, we perform the connectivity check as follows. On the client side every $T_{KA}$ [ms] we check if we have received at least a keep-alive response in the last $T_{TO}$ = K·$T_{KA}$ [ms]. If not, we declare the tunnel down. $T_{TO}$ is the "tunnel time out" interval and it is directly related to the "responsiveness" of the connectivity check procedure, that we define as the time $T_R$ needed to detect a faulty tunnel. As a measure of the responsiveness, we can use either the worst case delay $T_{Rmax}$ for declaring a tunnel down after a fault, or the average delay $T_{Ravg}$, as defined in eq. (1) and (2) (see Appendix II).

$$T_{Rmax}=(K+1)T_{KA}+ RTT_{max} = T_{TO}+T_{KA}+RTT_{max} \quad (1)$$

$$T_{Ravg}= (K+1/2)T_{KA}+ RTT_{avg}/2= = T_{TO}+T_{KA}/2+RTT_{avg}/2 \quad (2)$$

Even if the tunnel is active some consecutive keep-alive packets could be lost and this could lead to declaring the tunnel down (leading to a "false positive" event). Every time that a probe is sent, there is the probability $p_{fp}$ of declaring a tunnel down when it is still alive. This is equal to the probability of having K probe requests not acknowledged by the server-end; a probe request could be not acknowledged because either the probe request has been lost (which happens with probability $p_{loss-req}$) or the probe response has been lost (which happens with probability $p_{loss-res}$).

For the sake of simplicity, let us assume that max RTT < $T_{KA}$, that is, the maximum round trip time RTT is smaller than the keep alive period $T_{KA}$. Under this assumption, we can detect the loss of the probe request or of the probe reply $T_{KA}$ ms after sending the probe request: either the probe request has been received or a loss event has happened, because it is not possible that the probe response is delayed more than RTT < $T_{KA}$.

Let also assume that OWLc=OWLs=$p_{loss}$, i.e. the loss probability of the channel between the client-end and the server-end is the same in both directions (the analysis can be easily extended to the case where OWLc≠OWLs):

$p_{loss-req} = p_{loss}$

$$p_{loss-resp} = (1 - p_{loss}) * p_{loss}$$
$$p_{fp} = (p_{loss-req} + p_{loss-resp})^K = (2p_{loss} - p_{loss}^2)^K \quad (3)$$

Eq. (3) relates the tunnel loss probability $p_{loss}$ with the false alarm probability $p_{fp}$ for different values of K, the number of consecutive probes that need to be lost before declaring the tunnel down ($T_{TO} = K\, T_{KA}$). Obviously for a given K, $p_{fp}$ increases with $p_{loss}$. The false alarm probability $p_{fp}$ is not suited to be directly used as performance metric of the connectivity check procedure as the perceived impairment is proportional to the frequency of false positive events $F_{fp}$:
$$F_{fp} = p_{fp} * 1/T_{KA}$$
Therefore, we consider the reciprocal of $F_{fp}$, i.e. the average time $T_{fp}$ between two false positive events as the main performance metric of the connectivity check procedure:
$$T_{fp} = 1/F_{fp} = T_{KA}/p_{fp} = T_{KA}/(2p_{loss} - p_{loss}^2)^K \quad (4)$$
The resource consumption (CPU processing and network capacity) of the connectivity check procedure is directly proportional to the frequency of keep alive packets; therefore, the probe interval $T_{KA}$ should be as large as possible. In order to have a good responsiveness $T_{Ravg}$ should be as small as possible; in order to limit the impairments due to false positive events, $T_{fp}$ should be as large as possible. NB: we choose to refer to the average case using $T_{Ravg}$ modeled by eq. (2), but it would be possible to consider $T_{Rmax}$ and the worst case modeled by eq. (1), with very similar results.

By fixing a maximum keep alive rate (i.e. a minimum $T_{KA} = T_{KA}^{min}$) we can consider the tradeoff between responsiveness $T_{Ravg}$ eq. (2) and the average interval between false positive events in declaring a tunnel down $T_{fp}$ eq. (4) for different values of K. We also need to provide estimates of round trip time $RTT_{avg}$ and loss probability $p_{loss}$. As shown in Fig. 7 ($T_{KA}^{min}$ = 100 ms, $RTT_{avg}$ = 100 ms, $p_{loss}$ < 5%), by increasing K we have a linear increase of $T_{Ravg}$ (which corresponds to a worsening of the responsiveness) and an exponential increase of $T_{fp}$ (which means an improvement of the performance)

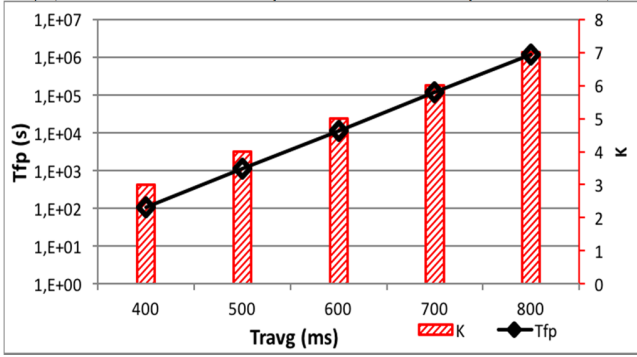

Fig. 7 Tradeoff between $T_{Ravg}^{max}$ and $T_{fp}^{min}$ ($T_{KA}$ and RTT fixed)

On the other hand, if we require the responsiveness $T_{Ravg}$ to be smaller than a target $T_{Ravg}^{max}$ and that average time between two false positive events $T_{fp}$ to be longer than a target $T_{fp}^{min}$, we can define the following optimization problem for $T_{KA}$:
$$\max T_{KA} \mid \begin{cases} T_{Ravg} < T_{Ravg}^{max} \\ T_{fp}^{min} < T_{fp} \end{cases} \quad (5)$$
The maximization problem (5) can be rewritten as:
$$\max T_{KA} \mid \begin{cases} (K + 1/2)\, T_{KA} + RTT_{avg}/2 < T_{Ravg}^{max} \\ T_{fp}^{min} < T_{KA}/p_{fp} \end{cases}$$
$$\quad (6)$$

$$\max T_{KA} \mid \begin{cases} T_{KA} < (T_{Ravg}^{max} - RTT_{avg}/2)/(K + 1/2) \\ T_{fp}^{min} * (2p_{loss} - p_{loss}^2)^K < T_{KA} \end{cases} \quad (7)$$

$T_{KA}$ is constrained by two inequalities: (6) is related to the responsiveness and (7) to the interval between false positive events in considering a tunnel down. Assuming a given maximum loss probability $p_{loss}$ and average round trip time $RTT_{avg}$, the combination of eq. (6) and (7) provides the admissible range for $T_{KA}$ depending on K. For example, let us assume $RTT_{avg}$ = 100 ms, $p_{loss}$ < 5%, $T_{Ravg}^{max}$ = 500 ms, $T_{fp}^{min}$ = $10^5$ s ($\approx$ 27,8 h). Fig. 8 plots the eq. (6) and (7) respectively labeled "Resp." and "FPev." and displays the admissible range for $T_{KA}$.

The optimal $T_{KA}^*$ constrained the by target performance parameters $T_{fp}^{min}$ and $T_{Ravg}^{max}$ can be found combining eq. (6) and (7) into
$$T_{fp}^{min} \cdot (2p_{loss} - p_{loss}^2)^K < \left(T_{Ravg}^{max} - \frac{RTT_{avg}}{2}\right) / \left(K + \frac{1}{2}\right) \quad (8)$$
Let $K^*$ be the minimum value for which the inequality (8) holds, we can chose the optimal $T_{KA}$ by using eq. (6):
$$T_{KA}^* = (T_{Ravg}^{max} - RTT_{avg}/2)/(K^* + 1/2)$$
Looking at Fig. 8, $K^*$ is the smallest K for which the FPev. curve goes below the Resp. curve, while $T_{KA}^*$ is the value of the Resp. curve for K= $K^*$. In our example, $K^*$ =7 and the $T_{KA}^*$ = 60 ms. The graphs in Fig. 9, Fig. 10 and Fig. 11 report the optimal K and $T_{KA}$ by varying respectively $p_{loss}$ (from 0.5% to 8%), $T_{fp}^{min}$ (from 100 s to $10^6$ s, which corresponds to ~11.5 days) and $T_{Ravg}^{max}$ (from 80 ms to 2 s), keeping all the other parameters as in the previous example.

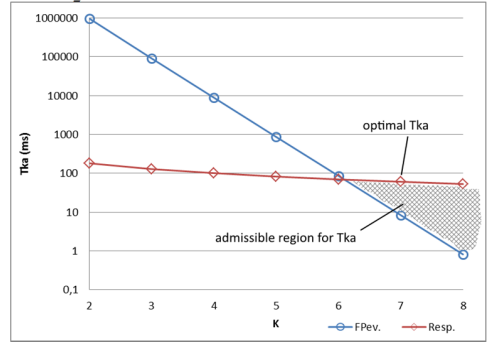

Fig. 8 Evaluation of $K^*$ and $T_{KA}^*$

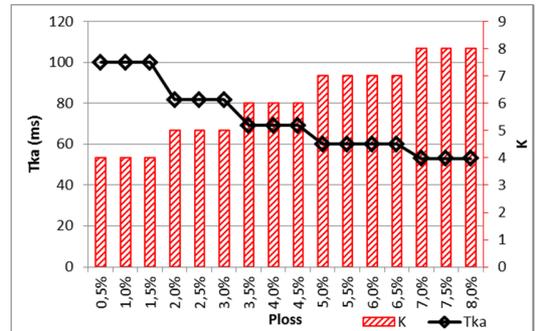

Fig. 9 $K^*$ and $T_{KA}^*$ vs. $p_{loss}$

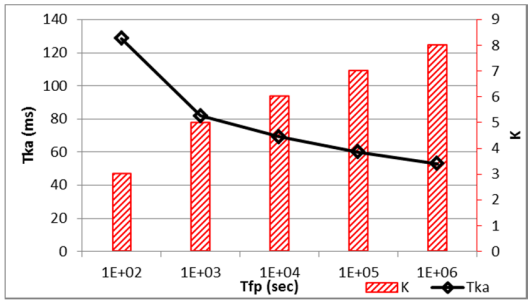

Fig. 10 $K^*$ and $T_{KA}^*$ vs. $T_{fp}^{min}$

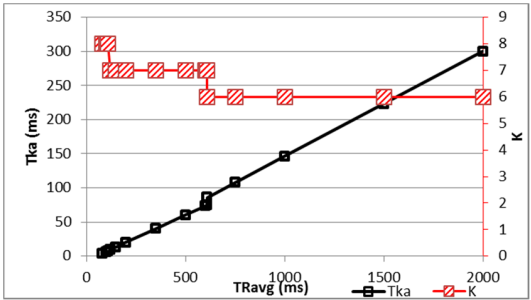

Fig. 11 $K^*$ and $T_{KA}^*$ vs. $T_{Ravg}^{max}$

## 5 IMPLEMENTATION DETAILS AND PERFORMANCE ASPECTS

The UPMT software is composed of a kernel module dealing with encapsulation of packets into tunnels and of a Java application that offers a GUI to the user and manages the signaling messages between the UPMT remote entities. The signaling, including the setup of the UPMT tunnels, is based on the SIP protocol and implemented using the Open Source MjSip stack [15].

The Keep-Alive procedures described in the previous sections have been implemented using different approaches: 1) a user space implementation leveraging the SIP protocol already used for tunnel setup signaling to carry the Keep-Alive information; 2) a kernel space implementation extending the UPMT tunneling module; 3) a more efficient user space solution in Java based on a UDP packet; 4) a reference stand-alone user space implementation in C also based on UDP. In [16] we have compared the performances of the SIP user space solution and of the kernel space solution in terms of processing cost. Here, we also consider the performance of the user space implementations based on UDP packets.

We emphasize that the source code of our implementation is available at [17]. On the UPMT project home page [6], we also provide a ready-to-go Virtual Machine to make our experiments more easily replicable.

### 5.1 User Space Implementations

In the SIP based implementation, the Keep-Alive probe packets are realized using SIP MESSAGE methods [18], a type of SIP request messages that do not create a session, but can be used to transfer any information. The receiving entity replies with a SIP 200 OK message according to the SIP protocol rules. The SIP protocol implementation manages multiple retransmission of the request if no reply comes in within a timeout. We enhanced the SIP stack adding a new SIP header to the messages, called Timestamp. When performing the Keep-alive procedure, the Mobile Host will send a SIP MESSAGE toward the correspondent UPMT node, adding the Timestamp header (time is expressed in millisecond since Jan 1 1970). The initial part of the SIP MESSAGE is reported in Fig. 12, showing the new Timestamp header. This solution was easy to implement because we reused functionality available in the SIP stack, but it suffers from poor performance.

The Java UDP based user space solution avoids the processing overhead introduced by the SIP protocol. The probe packet is a UDP packet encapsulated within the tunnel (Fig. 13-A). The external IP destination address and UDP destination port are the ones of the tunnel. The internal IP destination address is the same of the tunnel, specific UDP source and destination ports are used to distinguish Keep-Alive probe packets from regular UDP packets. In our implementation, we have reserved these ports so that they are never allocated to UDP sockets.

As we will discuss in section 5.4, the performance of this Java UDP based solution in terms of processing load is still too poor compared to the kernel based solution. Therefore we decided to prepare a reference user space implementation of the Keepalive procedures in C, operating as a stand-alone client server application (i.e. not integrated in the UPMT implementation). We refer to this application as *karle* (Keep Alive with Rtt and Loss Estimation). *Karle* is a minimalistic single threaded application that executes the Keep alive procedures between a single client and a single server, reading and writing on a UDP socket. It does not handle the monitoring of multiple connections in parallel, keeping only the state information for a single connection. As such, it provides an upper bound in terms of packet processing capacity of an application capable of handling multiple connections. Being available as open source [7], *karle* code base can be used as a library to be integrated in other mobility management solutions.

```
MESSAGE sip:160.80.103.66:5060 SIP/2.0
Via: SIP/2.0/UDP 5.6.7.8:40000;rport;branch=z9hG4bK809f
Max-Forwards: 70
To: <sip:160.80.103.66:5060>
From: <sip:1.2.3.30>;tag=251807832719
Call-ID: 314335872631@5.6.7.8
CSeq: 1 MESSAGE
Expires: 3600
User-Agent: mjsip 1.7
Timestamp: 1339598185957
```
Fig. 12 SIP MESSAGE for the Keep alive probe

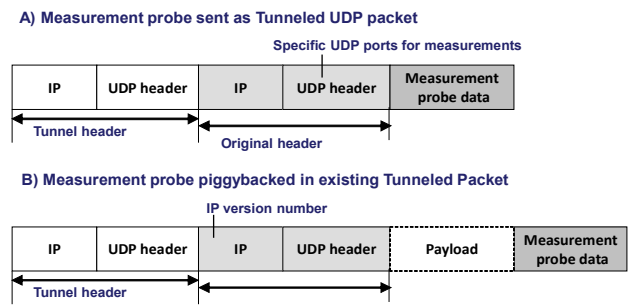

Fig. 13 Probe packet formats

The format of the measurement probe data is shown in Fig. 14. It allows to perform the RTT and loss evaluation in a combined way, three 32 bits words are used for the RTT and three 32 bits words for the loss evaluation (the total is 24 bytes). Even if not used in the algorithms, in the implementation we also number all sent packets with a 32 bit sequence number, so that the actual number of sent bytes is 28.

| RTT evaluation | | | LOSS evaluation | | |
|---|---|---|---|---|---|
| 32 bit | 32 bit | 32 bit | 32 bit | 32 bit | 32 bit |
| $t_{Sc}$ | $t_{Ss}$ | $\Delta t$ | Sc | Ss | R |

Fig. 14 Content of measurement probe data

### 5.2 Kernel Space Implementation

In the kernel space implementation, the Keep-Alive procedures with RTT and loss ratio evaluation are performed within the UPMT Linux kernel module. Linux kernel timers are used to schedule the sending of probe packets for each active tunnel. It is possible to activate/deactivate the Keep-Alive procedures for each tunnel by sending configuration commands from a user space application.

We implemented the algorithms described in section 4 and used the packet format of Fig. 13-A. When originating a probe packet, the kernel module encapsulates the inner probe packet into a UDP packet and sends it. When receiving a probe packet, the kernel module decapsulates the packet like any other packet received on the tunnel. Then a matching with UDP destination and source ports is performed to recognize the probe packets. If the packet is recognized as a probe, it will not be forwarded to a UDP socket to be delivered to user space but it will be analyzed in the kernel. In this case, the kernel module generates the probe reply packet (copying the timestamp from the received packet) and encapsulates it into a UDP packet to be sent back to the sender of the probe. We have further improved this solution with the possibility to piggyback the measurement information inside data packets, as described in the next subsection.

The current UMPT kernel module provides the evaluated RTT(k) and OWL(k) to the Java user space application. The EWMA algorithm (whose details are discussed in the Appendix I) is performed by the Java user space application. This leaves a further optimization margin as the EWMA could be moved in kernel space. In this case, the arbitrary exponentiation operations in eq. (13) needs to be properly replaced by multiplications and divisions considering that floating point operations in kernel are discouraged or not allowed by kernel configuration.

### 5.3 Optimization with Piggybacking

The Keep-Alive procedures described in section 4 are based on sending periodic probes with period $T_{KA}$ (*active* approach). Assuming that it is possible to piggyback keep-alive information on existing packets, we propose an improved mechanism, referred to as *active-pb*. The basic idea in the *active-pb* approach is to ensure that for each $T_{KA}$ period a probe request (or probe response) is sent by the client-end (or by the server-end). The packet tunneling module in the kernel tries to piggyback the probe information in existing packets during the $T_{KA}$ time interval. If it is not possible to piggyback the probe information during the $T_{KA}$ interval, an active probe packet is sent at the $T_{KA}$ expiration.

Fig. 15 illustrates the *active-pb* approach. Note that the client and server clocks do not need to be synchronized because they can measure the $T_{KA}$ interval independently. The maximum time interval between two probe requests is $2*T_{KA}$ and the maximum delay introduced by server due to the passive piggybacking attempt is $T_{KA}$. The same RTT and OWL evaluation procedures described for the *active* approach can be reused in the *active-pb* approach.

In the *active-pb* approach, time stamps and packet counters information are added to packets in transit on a tunnel. This information is added only to IP packets with a length shorter than a threshold, so that the addition will not cause the packet to exceed the Maximum Transmission Unit (MTU) of crossed links. Note that this optimization is only possible when the mobility management solution is based on some form of tunneling that can be enhanced with this mechanism. In our implementation, the information is piggybacked and extracted by the UPMT tunneling module while encapsulating and de-capsulating the packets in the tunnel (this is performed in kernel space with a minimal CPU overhead). The packet format used for the piggybacked packets is shown in Fig. 13-B. The measurement data are added at the end of the packet. In the first byte of IP header the *version* field is normally used to indicate the IP version used in the packet. Since this field is 4 bit length, we set the value of 15 (all the bit are set to 1) in case of piggybacked packet. In the packet receiving procedure we use the version field of the inner IP header (the original header) to check if the current packet is piggybacked or not. If yes, we restore the normal value of this field (the number 4 for IPv4 packets), we remove the measurement information and the packet is sent to the upper levels of the networking stack. Clearly this is possible because our UPMT tunnels are only meant for IPv4 packets (the approach can be easily extended to support IPv6 packets in the tunnel).

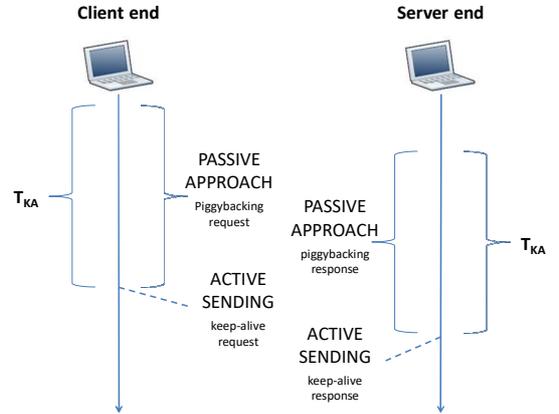

Fig. 15 Active-pb approach

The *active-pb* approach can achieve a saving of the capacity, especially in wireless technologies which are not efficient in sending small packets. For example in 802.11b the air time for sending 24 bytes of data (corresponding to the Keep Alive payload) is in the order of 800 µs (including an average backoff and assuming no collisions), of which only 17 µs are used for the 24 bytes of payload. Piggybacking the keep alive information saves 98% of the additional capacity needed for the Keep Alive procedures.

### 5.4 Processing Performance

Setting the Keep alive rate at the highest possible value allows to have a more precise estimation of RTT and of Round

Trip loss ratio and to react in a faster way to changing network conditions. Unfortunately, there are two factors that limit the increase of the Keep alive rate: the CPU load on the Mobile Hosts and intermediate mobility management nodes, if present and the network load. Of these two factors, the CPU load is the most critical one since in both the *Internet access* and the *Peer-to-peer multi-access* scenarios it affects the battery usage. Even if the CPU load due to the monitoring of few tunnels would be low in absolute terms, a reduction of this load has a positive impact on battery duration, as the performance monitoring procedure needs to be continuously executed when the Mobile Host is connected. Considering a mobility management node (i.e. the Anchor Node) in the *Internet access* scenario, the CPU processing due to the monitoring procedures can even be the bottleneck for the node. As a rule of the thumb, a Keep alive rate in the order of 2-3 Keep alive per second could be enough to fulfill the requirements of a precise and timely estimation of RTT & loss and of a timely detection of connectivity losses. From the network load perspective, this would correspond to few hundred bit/s, i.e. one order of magnitude less than a VoIP call. On the other hand, we show hereafter that the CPU load may become critical even at these relatively low rates.

We set up our testbed with virtual machines running on VirtualBox [19] in a PC with an Intel® Core™2 Quad CPU Q8400 processor running at 2.66Ghz (4GB RAM). We focused on the *Internet access* scenario and considered the CPU utilization in the Anchor Node. One virtual machine was acting as an Anchor node, while the Mobile Hosts were running in different virtual machines. We executed the Keep alive procedures at different rates both for the user space and for the kernel space implementations. We measured the CPU utilization using the *sar* command, a part of the *sysstat* package. Further results and more details on the experiments can be found in [13][16]. The CPU utilization grows linearly with the sending rate of the probes. We were able to estimate the maximum Keep alive rate within a given CPU utilization threshold (e.g. 50%) for the Anchor node, reported in Table 2.

For the SIP-based user space implementation, the maximum keep-alive rate in our experiments is 100 (s$^{-1}$). The UDP based user space java implementation improves this performance by almost a factor of 3. In the same conditions the *karle* C application is able to manage a keep-alive rate of around 3000 (s$^{-1}$) at 50% CPU load. The kernel space implementation is one order of magnitude more efficient than the user space *karle* application.

If we assume 2 Keep-alive per second per tunnel and 2 tunnels per client the maximum number of clients for a mobility management node corresponds to the maximum keep alive rate reported in Table 2 divided by 4. This would roughly lead to a number of 25, 70, 750 or 11.600 supported users for the four different implementations. Clearly, these results are dependent on the specific hardware that we have used for the experiment, but what is of general interest is the empirical evaluation of the ratio between the supported number of flows in the user space solutions and the one in the kernel space solution.

Table 2 Maximum Keep-alive rate for 50% CPU utilization

|  | SIP | UDP Java | UDP C *karle* | Kernel |
|---|---|---|---|---|
| Max rate (msg/s) | 100 | 278 | 3000 | 46512 |

As we mentioned above, this is an important indication also for the CPU processing load in the Mobile Host side, which we have not explicitly measured. Such a large reduction of the processing load for the kernel-based solution has a benefic impact on the battery duration.

## 6 ESTIMATION ACCURACY AND TIMELINESS

In this section, we discuss the accuracy and timeliness of the methodologies described in section 4 for the monitoring of round trip time (RTT). Additional details and some results related to the evaluation of loss ratio (OWL) are reported in Appendix IV. We provide design guidelines to tune the parameters of the proposed mechanisms. We evaluate how much the provided measurements are close to the real network conditions and how the algorithms react to the variation of network quality (delay and loss). The accuracy of the measurements is very important as it is used to drive the handover procedures: a bad estimation of the RTT and loss probability would lead to sub-optimal handover decisions, impairing the QoE perceived by the user.

The RTT and OWL measurement samples are accumulated using the generalized EWMA algorithm described in eq. (13) of Appendix I. For the delay estimation (RTT) the samples are available on average every $T_{KA}$ seconds (assuming that there are no losses of probe packets). For the loss measurement (OWL) the samples are available every $T_L = N \cdot T_{KA}$ seconds.

The *time constant* $\tau$ of the generalized EWMA, defined in Appendix I, determines how the measured samples of RTT and OWL are averaged over time; the choice of the appropriate value for $\tau$ is a critical design choice. A longer $\tau$ provides an average over a longer period of time but it makes the EWMA slower to react to changes. A shorter $\tau$ makes the system more responsive to changes but it includes in the EWMA only the more recent measurements.

We can relate the time constant to the responsiveness/timeliness of the estimator, $T_{tml}$ as the time required for the RTT/OWL estimate to fall within an interval smaller than 10% of the RTT/OWL variation when a change occur. As shown in Appendix IV, we have that:

$$T_{tml} = \tau \log 10 \approx 2.3\,\tau$$

Faster detection implies smaller values of $T_{tml}$ which translates to smaller value of $\tau$. For instance, a timeliness of $T_{tml}$=2sec, results into a time constant $\tau \approx 0,87$sec. Hereafter, since we are interested in the measurements we will refer to the time constant $\tau$ rather than to the timeliness $T_{tml}$.

We have performed a set of experiments over the final version of our prototype implementation (described in section 5.2) and collected the measurements results reported in this section. In all the experiments we use two UPMT hosts connected though a Linux PC acting as a router. We used the *netem* [20][21] module of the Linux kernel in the Linux router to generate tunable delay and loss ratio on the outgoing interfaces. In the first experiment (RTT step variation) we start with an RTT delay of 100 ms and then we sudden increase the delay to 200 ms (this happens at time t=3.75 s in Fig. 16). Fig. 16 plots the EWMA estimate of the RTT compared to the thin dotted line that represents the reference RTT (i.e. the RTT that we have imposed on the path) using four different time constants. The figures also report the duration of the time constants and of the $T_{KA}$ interval in scale with the x axis. In Fig. 16 the keep alive procedure for RTT measurement has $T_{KA}$ = 200 ms, the four values of the time constant τ are 124, 218, 392 and 896 ms. We used eq. (13) with reference time interval T=$T_{KA}$, and four decreasing values of α (0.8, 0.6, 0.4, 0.2) to obtain the reported time constants. As expected, using relatively small time constants the EWMA quickly follows the step variation of the RTT, but the EWMA significant contribution is coming only from the last 2 or 3 measurement values.

The choice of the time constant depends on the variability of the RTT and on the dynamicity of the control decisions that can be taken based on the measured RTT. If it is possible to react in the order of seconds, the time constant should be small enough to measure the performance of the last seconds, but such small time constants are not useful if the reactions/decisions are taken in the order of tens of seconds.

In the second experiment (RTT 3-levels, Fig. 17), we create an RTT with a periodic behavior. It has a period of 20 seconds, in which it alternates among 3 levels: 200 ms for 5 s, 300 ms for 5 s, 200 ms for 5 s, 100 ms for 5 s. In Fig. 17, $T_{KA}$ = 2 s , the EWMA is plotted for time constants τ of 1.24 and 8.96 s. It can be seen here that with a time constant of 8.96 s it is not possible to track the variations of RTT, which changes every 5 s and the resulting EWMA filters out the maximum and minimum values of RTT, oscillating around the average.

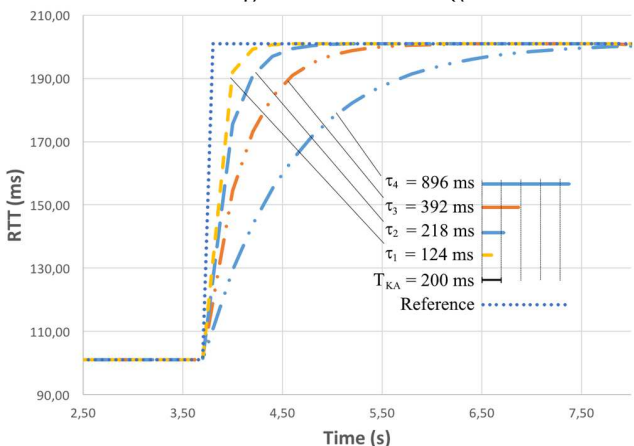

Fig. 16 - RTT step variation ($T_{KA}$ = 200 ms), different time constants

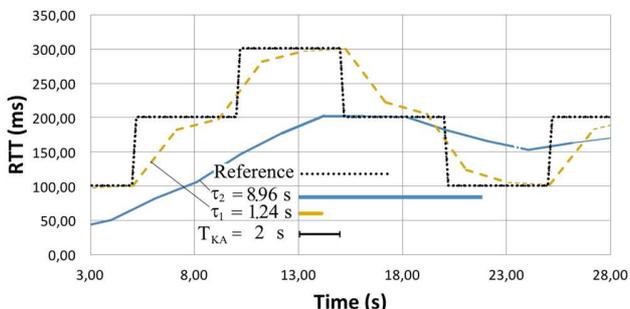

Fig. 17 - RTT 3-levels ($T_{KA}$ = 2 s) with different time constants

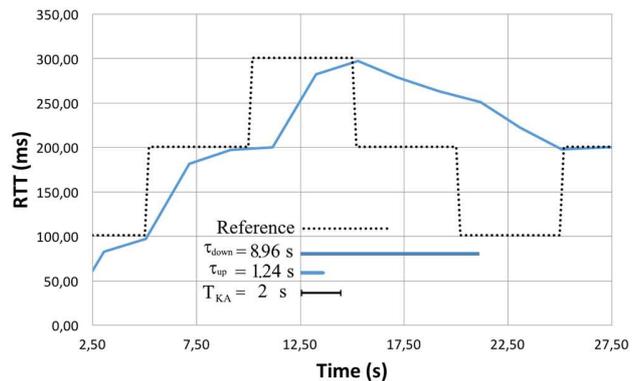

Fig. 18 - Asymmetric RTT approach, $T_{KA}$ = 2 s

By looking at Fig. 17, we realized that in some scenarios it is not bad that the EWMA does not decrease too quickly. In particular, when a parameter like the RTT decreases for a short interval and then it increases again it could be misleading that the EWMA algorithm reports the improvement, only to counteract few seconds after and report a new increase of the delay. We think that a shorter time constant can be used to process an RTT sample that reports a worsening of the network performances (i.e. an increase of the RTT), while a longer time constant can be used to process the samples that report an improvement (i.e. a decrease of the RTT). Using this new approach in the experiment (RTT 3-levels) reported above, we obtain the results shown in Fig. 18. It can be seen that the transitions with an increase of RTT are followed promptly, while the transitions with a decrease of the RTT are followed more slowly. In particular, it can be seen from Fig. 18 (a scenario in which the EWMA estimation cannot accurately follow the RTT variation) that the EWMA of the RTT that is reported in this case is more realistic as it follows more closely the higher RTT delays. In fact, in case of an RTT oscillation more frequent than our capability to capture it, it is much better to report a value close to the maximum of the RTT in the period rather than its average value.

## 7 RELATED WORK

Accurate performance monitoring mechanisms at one or more of the protocol layers are fundamental to any mobility management system over heterogeneous networks and several articles and solutions have been proposed.

As previously observed, the physical layer measurements, e.g., the Receiver Signal Strength, while providing key information on the wireless link status, do not reflect the corresponding end-to-end performance. Link layer statistics, as the Transmission Error Detector proposed in [22] provide the upper layers with information about the reception of packets at the wireless AP to drive interface selection. They show good performance, but are only effective when the bottleneck is represented by the wireless access links.

At the network layer, a number of metrics can be measured, from the simpler ones like bandwidth, delay and packet loss to more complex ones like network reliability, security, cost, and load. Most approaches (see [3] and references therein) focus on the architectural and protocol aspects of mobility management and provide little or no details on the underlying measurements procedure. For instance, in [23] the authors consider an architecture for network mobility within the context of the IETF NEMO (Network Mobility) WG and only mention that "There are a number of metrics that can be measured, with the most common being QoS metrics such as bandwidth, delay and packet loss". Other approaches exclusively rely on measurements taken on user traffic. In [24], to setup a connection to a peer, a node sends multiple SYN packets, one for each interface. The node only completes the three-way handshake with the first received SYN-ACK packet thus implicitly selecting the interface with the current smallest RTT. After the connection has been setup, no further monitoring is considered.

Similar in scope to our work, the WiOptiMo [25] solution aims to provide seamless and continuous connectivity by adaptively selecting the best internet connection among all wireless access providers available, guaranteeing persistence in case of signal and/or performance degradation. To this end, ICMP control packets (ping) are periodically sent to the access points to estimate the connection performance on the basis of the experienced Round-Trip-Time (RTT). The link disconnections are detected whenever 3 consecutive ping packets are not responded within a predefined timeout (based on empirical evidence that the failure of a connection can be robustly assessed after three consecutive timeouts [26]).

In the context of the SHIM6 approach to Multi-homing [27], the Failure Detection and Locator Path Exploration Protocol (REAchability Protocol, REAP for short) [28] has been designed to detect failures in the currently used path and to identify a new path. The protocol relies on two timers driven by upper layers traffic and by the so called Keep Alive messages when traffic is sporadic, to detect failures. The Keep Alive timer is started whenever a data packet is received and stopped and reset each time the node sends a data packet to the peer. A Keep Alive message is sent whenever the Keep Alive timer expires. The Send Time timer is started when the peer send a packet and stopped whenever a packet is received. When the Send Timer expires, a fault is detected. The specification [28] suggests to use a Send Timer timeout larger than 10 seconds to avoid reacting to temporary failures. This high timeout value clearly impacts the fault detection efficacy. In [29] the authors show how the Send Timer greatly affects the traffic recovery time in case of failures, with a simple simulation approach. In [30] the authors propose an analytical model of REAP and derive analytical expression for the recovery time from a path failure as seen by the upper layers in different scenarios. The work in [31] does not focus on the performance of the REAP recovery procedure seen by the host, but on the scalability of the solution in large scale network.

Differently from REAP, our proposed approach caters for temporary failures and/or packet loss by requiring the loss of multiple packets before declaring a link down. As shown in Section 4.3, this allows us to trade-off the detection responsiveness with the probability of a false positive by adjusting the value of the $T_{KA}$ and K parameters. Another difference is that REAP monitors the active path and starts a recovery procedure if it fails, while in our approach we can monitor the active and alternate paths (tunnels) in parallel to perform a seamless handover when possible.

RTT and one way delay estimation have been also investigated in the literature in the broader context of network performance measurements. In [32] the author compares the implementation of three RTT estimation algorithm: Jacobson's algorithm based on EWMA and commonly used by TCP, Expert Framework and Eifel algorithm. The last two are examined more deeply in [33] and [34] respectively, and they are considered as a starting point to build complex solution about RTT estimation. In [35], the authors analyze EWMA parameters in TCP retransmission timeout estimation.

In [36] and [37], the authors consider techniques to monitor the One Way Delay in a passive way with minimal overhead. A similar approach is used in [38], dealing with passive RTT measurement. In our scenario, such solutions would require sending lists of packet hashes and timestamps over the wire and additional processing load to compute hashes and to search matches among the packet hash lists.

Some works consider the performance monitoring from the perspective of overall network management rather than from the perspective of mobility management/ vertical handover for Mobile Hosts. For example in [39] a complete tool for evaluation of network performance in terms of various metrics is described, while in [40] the focus is given to RTT for TCP flows.

Multipath TCP (MPTCP, [42]) is a solution for exploiting multiple network paths among to two end-points. MPTCP is meant for TCP based applications and it is a complete solution offered to the application. In the course of its operations, MPTCP needs to take measurements of the network performance. Our proposed Keep Alive procedures work at IP level and are meant to support all types of applications, providing a tool that needs to be integrated in a mobility management architecture.

In order to overcome the limitation of performing networking related operation in user space, frameworks for fast packet I/O have been proposed, see for example [43]. A direction for future work is to design and implement the Keep Alive procedures within such frameworks.

Finally, we mention the ITU-T recommendation on operation and management for Ethernet [41]. It describes a "continuity check" procedure for protection switching and a set of functions for performance monitoring, in particular frame loss ratio, whose requirements are very similar to our needs.

## 8 CONCLUSIONS

In this paper, we have presented novel solutions for the performance monitoring of wireless access network interfaces to support the handover decision process. Our solution include computationally and memory efficient procedures for the timely estimation of the Round Trip Time and of Round Trip and One Way loss ratio.

We have proposed a connectivity check procedure, addressing the important issue of the trade-off between responsiveness and false alarm probability and proposing an analytic approach to find the optimal setting of the parameters.

We have implemented the proposed solutions on the Linux OS in user space and in kernel space and performed comparative measurements of CPU utilization. As expected, the kernel space solution is more efficient in terms of computational load and thus energy consumption. Consequently, the kernel version is the implementation of choice both for a mobility management node that is expected to handle thousands of Mobile Hosts concurrently and for the Mobile host where battery duration is the main concern. As an alternative, the fast I/O frameworks or the so called *kernel* *bypass* solutions should be considered to avoid the bottleneck of user-to-kernel communication.

The source code of our implementation is publicly available. For an easier reproducibility of the results, we have also provided a ready-to-go Virtual Machine with scripts and instructions for the setup of the experiments.

## 9 ACKNOWLEDGEMENTS

The authors wish to thank Marco Bonola for his work on the design of UPMT and his precious suggestions and Marco Galvagno for the implementation of the Java user space monitoring solution.

## 10 REFERENCES

[1] D. Le, X. Fu, D. Hogrere, "A Review of Mobility Support Paradigms for the Internet", IEEE Communications surveys, 1s t quarter 2006, Volume 8, No. 1

[2] A. Gladisch, R. Daher, D. Tavangarian, "Survey on mobility and multihoming in future internet", Wireless personal communications, 74(1), 2014

[3] S. Ferretti, V. Ghini, F. Panzieri, "A survey on handover management in mobility architectures", Computer Networks, 94, 2016

[4] Meriem Kassar, Brigitte Kervella, Guy Pujolle, "An overview of vertical handover decision strategies in heterogeneous wireless networks", Computer Communications, 31(10), 2008, Pages 2607–2620

[5] Les Cottrell, "Network Monitoring Tools", available on line at http://www.slac.stanford.edu/xorg/nmtf/nmtf-tools.html

[6] UPMT homepage: http://netgroup.uniroma2.it/UPMT

[7] Karle application – "Keep Alive with Rtt and Loss Evaluation", https://github.com/netgroup/karle.git

[8] M. Bonola, S. Salsano. "UPMT: Universal Per-Application Mobility Management using Tunnels", IEEE GLOBECOM 2009

[9] C. Perkins, Ed., "IP Mobility Support for IPv4, Revised", IETF RFC 5944, November 2010

[10] R. Moskowitz, P. Nikander, T. Henderson, "Host Identity Protocol", IETF RFC 5201, April 2008

[11] Distributed Mobility Management, IETF Working Group home page http://datatracker.ietf.org/wg/dmm/

[12] H. Chan et al., "Requirements for Distributed Mobility Management," IETF RFC 7333, Apr. 2014.

[13] S. Salsano, M. Bonola, F. Patriarca, "The UPMT solution (Universal Per-application Mobility Management using Tunnels)", technical report available at http://netgroup.uniroma2.it/TR/UPMT.pdf

[14] M. Mathis, et al. "The macroscopic behavior of the TCP congestion avoidance algorithm", ACM SIGCOMM Computer Communication Review 27.3, 1997.

[15] MjSip home page: http://www.mjsip.org

[16] F. Patriarca, S. Salsano, F. Fedi, "Efficient Measurements of IP Level Performance to Drive Interface Selection in Heterogeneous Wireless Networks", PE-WASUN'12, October 21 – 25 2012, Paphos, AA, Cyprus

[17] UPMT source code, https://github.com/StefanoSalsano/UPMT

[18] B. Campbell (Editor),"Session Initiation Protocol (SIP) Extension for Instant Messaging", IETF RFC 3428, December 2002

[19] Oracle VM VirtualBox, http://www.virtualbox.org

[20] Netem module in Linux kernel http://www.linuxfoundation.org/collaborate/workgroups/networking/netem

[21] S. Hemminger, "Network Emulation with NetEm", Linux Conf Au 2005, Camberra, Australia, 2005

[22] V. Ghini, S. Ferretti, F. Panzieri, "The "Always Best Packet Switching" architecture for SIP-based mobile multimedia services", Journal of Systems and Software Vol. 84, Issue 11, Nov. 2011.

[23] I. Alsukayti, C. Edwards, "Multihomed Mobile Network Architecture", IFIP Networking Conference, Tolouse, May 20-22, 2015.

[24] T. Yamaguchi et al., "An Optimal Route Selection Mechanism for Outbound Connection on IPv6 Site Multihoming Environment", IEEE COMPSACW 2013, Japan.

[25] G. A. Di Caro, et al., "A Cross-Layering and Autonomic Approach to Optimized Seamless Handover", WONS 2006, Les Ménuires (France), 2006.

[26] H. Velayos and G. Karlsson, "Techniques to Reduce IEEE 802.11b MAC Layer Handover Time", Vol. 3, TRITA-IMIT-LCN, KTH, Sweden, 2003.

[27] E. Nordmark, M. Bagnulo, "Shim6: Level 3 Multihoming Shim Protocol for IPv6", IETF RFC 5533, June 2009.

[28] J. Arkko, I. van Beijnum, "Failure Detection and Locator Pair Exploration Protocol for IPv6 Multihoming", IETF RFC 5534, June 2009.

[29] A. de la Oliva, et al., "Performance analysis of the REAchability protocol for IPv6 multihoming", NEW2AN 2007, St. Petersburg, Russia, 2007

[30] A. De la Oliva, et al., "Analytical characterization of failure recovery in REAP", Computer Communications 33.4 (2010): 485-499

[31] H. Naderi, B. Carpenter, "A performance study on reachability protocol in large scale ipv6 networks", IEEE ICCNT 10, Karur, India, 2010.

[32] S. Lukin, "A Comparison of Round-Trip Time Estimation Algorithms", Loyola University Maryland, UCSC SURF-IT Research, 2010.

[33] B. A. Nunes, et al. "A Machine Learning Approach to End-to-End RTT Estimation and its Application to TCP". IEEE ICCCN 2011.

[34] R. Ludwig, K. Sklower, "The Eifel retransmission timer", ACM SIGCOMM Computer Communication Review, 2000.

[35] M. Allman, V. Paxson, "On estimating end-to-end network path properties", ACM SIGCOMM Computer Communication Review, 1999

[36] S. Zander, G. Carle, T. Zseby, "Evaluation of Building Blocks for Passive One-way-delay Measurements", Passive Active Measurement Workshop (PAM 2001), Amsterdam, Netherlands, 23-24 April 2001

[37] S. Niccolini, et al. "Desing and implementation of a one way delay passive measurement system", IEEE/IFIP NOMS 2004.

[38] S. Zander, G. Armitage, "Minimally Intrusive Round Trip Time Measurements Using Synthetic Packet-Pairs", IEEE LCN 2013.

[39] J. Prokkola, et al. "Measuring WCDMA and HSDPA Delay Characteristics with QoSMeT," IEEE ICC 2007, Glasgow, Scotland

[40] P. Romirer, et al. "Network-wide measurements of TCP RTT in 3G", TMA 2009, published in LNCS vol. 5537.

[41] ITU-T Recom. G.8013/Y.1731, "OAM functions and mechanisms for Ethernet based networks", 07/2011

[42] D. Wischik, C. Raiciu, A. Greenhalgh, M. Handley, "Design, Implementation and Evaluation of Congestion Control for Multipath TCP", in NSDI, vol. 11, pp. 8-8. 2011

[43] L. Rizzo, "Netmap: a novel framework for fast packet I/O", 21st USENIX Security Symposium (USENIX Security 12). 2012.

## 11 APPENDIX I: EWMA EVALUATION

The definition for the EWMA $S_k$ of a variable x available at regular time intervals $\{t_k\}$ with period T ($t_k = k \cdot T$) is:

$$S_k = \alpha \cdot x_k + (1-\alpha) \cdot S_{k-1}$$
$$S_0 = x_0 \qquad (9)$$

where $S_k$ is the EWMA of x at time $t_k = k \cdot T$, and $\alpha$ ($0<\alpha<1$) is the "smoothing factor". A higher $\alpha$ implies a higher weight of more recent observations of x. Instead of $\alpha$ we can use the time constant $\tau$ to characterize the EWMA computation. We define the time constant $\tau$ as the time needed for the EWMA to decay to 1/e of its initial value when all the new observations are 0. Let $x_0 = c > 0$, $x_k=0$ $\forall k>0$. Then, from (9):

$$S_k = c(1-\alpha)^k \qquad (10)$$

According to (10), the EWMA will exponentially decay to 0. To evaluate the time constant $\tau$, we first evaluate $\bar{k}$ (11) and then use it to evaluate $\tau$ in eq. (12):

$$c(1-\alpha)^{\bar{k}} = c/e \Rightarrow \bar{k}\ln(1-\alpha) = -1 \Rightarrow \bar{k} = -1/\ln(1-\alpha) \qquad (11)$$
$$\tau = \bar{k}T = -T/\ln(1-\alpha) \qquad (12)$$

We generalize the EWMA and $\tau$ definition for the case in which the values of the variable x are available at non-regular intervals. Let $\{t_k\}$ be the sequence of time instants at which an observation $x_k$ is available. Let $\Delta_{tk} = t_k - t_{k-1}$. Given a reference time interval T, we define the generalized EWMA with smoothing factor $\alpha$ and reference interval T as follows:

$$S_k = \left[1 - (1-\alpha)^{\Delta_{tk}/T}\right] \cdot x_k + (1-\alpha)^{\Delta_{tk}/T} \cdot S_{k-1}$$
$$S_0 = x_0 \qquad (13)$$

According to (13), the smoothing factor used to take into account a given observation $x_k$ into the EWMA $S_k$ now depends on the time elapsed from the previous observation. If $\Delta_{tk} = T$ the smoothing factor is exactly $\alpha$.

We can evaluate again the time constant from (13):

$$c(1-\alpha)^{\tau/T} = c/e \Rightarrow \tau/T \ln(1-\alpha) = -1 \Rightarrow \tau = -T/\ln(1-\alpha) \quad (14)$$

The reference time interval T and $\alpha$ in eq. (13) do not constitute two independent degrees of freedom, because the behavior of (13) only depends on the *time constant* $\tau = -T/\ln(1-\alpha)$. In fact, we can rewrite the factor in eq. (13) that depends on T and $\alpha$ as follows:

$$(1-\alpha)^{\Delta_{tk}/T} = (1-\alpha)^{\Delta_{tk}/-\tau\ln(1-\alpha)} =$$
$$= \left[(1-\alpha)^{-1/\ln(1-\alpha)}\right]^{\Delta_{tk}/\tau}$$

The term in square brackets can be evaluated:

$$(1-\alpha)^{-\frac{1}{\ln(1-\alpha)}} = e^{\ln\left((1-\alpha)^{-\frac{1}{\ln(1-\alpha)}}\right)} = e^{-\frac{1}{\ln(1-\alpha)}*\ln((1-\alpha))} =$$
$$= e^{-1}$$

Therefore

$$(1-\alpha)^{\Delta_{tk}/T} = e^{-\Delta_{tk}/\tau}$$

and eq. (13) can be rewritten as follow, only depending on the *time constant* $\tau$

$$S_k = \left[1 - e^{-\Delta_{tk}/\tau}\right] \cdot x_k + e^{-\Delta_{tk}/\tau} \cdot S_{k-1}$$

## 12 APPENDIX II: RESPONSIVENESS

In section 4.3, we used the delay before declaring a tunnel down after the actual loss of connectivity as a measure of the responsiveness of the connectivity check procedure. We defined the worst-case delay $T_{Rmax}$ and the average delay $T_{Ravg}$ and provided their expressions in terms of $T_{TO}$, $T_{KA}$, $RTT_{max}$ in eq. (1) and (2). Fig. 19 helps clarifying how these expressions have been derived. The timeout $T_{TO}$ before declaring the tunnel down is $T_{TO} = K \cdot T_{KA}$ [ms], as an example in Fig. 19 we let K=2. As shown in Fig. 19, in the worst case a fault can happen in the outgoing path immediately after the sending of a probe packet. The probe response can come back (after RTT ms) immediately after the sending of another probe packet. In this case, K+1 $T_{KA}$ time intervals are needed before declaring the tunnel down, as in eq. (1). The average case is derived in a straightforward way, assuming that the fault can happen with uniform probability in any point of the outgoing and incoming paths and that the last response probe can be received with uniform probability in any point of a $T_{KA}$ interval.

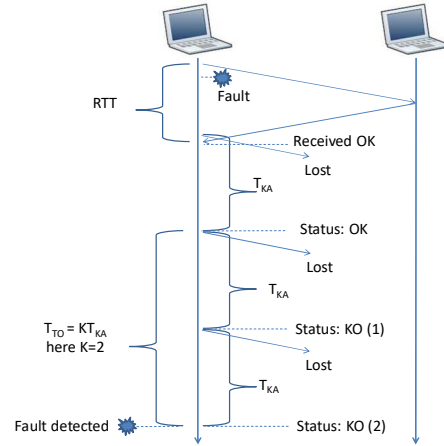

Fig. 19 Worst case delay before declaring a tunnel down

## 13 APPENDIX III: ROUND TRIP LOSS EVALUATION

Let us consider a generic request going from client to server, which expects a reply from the server. We can define as Round Trip Loss (RTL) ratio the fraction of lost replies with respect to the transmitted requests. The RTL takes into account that a packet can be dropped when travelling from the client-end to the server-end or in the way back from the server-end to the client-end.

RTL can be expressed as function of the "uplink" loss OWLc (from client-end to server-end) of the "downlink" loss OWLs (from server-end to client-end):
RTL = OWLc + OWLs – (OWLc * OWLs)

When OWLc and OWLs are small:
RTL ≈ OWLc + OWLs

We now describe a simplified RTL evaluation procedure that does not require the evaluation of OWLc and OWLs. Obviously, knowing RTL is suboptimal in case one needs a separate estimation of OWLc and OWLs but there is a saving in the complexity of the procedure and in the state information to be maintained.

The client-end evaluates the Round Trip Loss ratio over a time interval equal to $T_L = N \cdot T_{KA}$ where $T_{KA}$ is the configured interval for the Keep Alive procedure. Using the same notation of the previous section, we define as Sc and Rc the total number of probe requests sent and probe response received by the client-end on the tunnel. More precisely, the client-end increases the Sc variable for each probe packet sent and the Rc variable for each probe response received in the tunnel.

The RTL is evaluated on the client-end when receiving the first probe response after the $T_L$ expiration. For each RTL evaluation, the client-end sends the RTL value to the server using the first available probe request.

The sequence of evaluated RTL values will be denoted as RTL(m).
On the client-end:
RTL(m) = max[1 – ((Rc(k)–Rc_last)/(Sc(k)–Sc_last));0]
Sc_last ← Sc(k)
Rc_last ← Rc(k)–max[((Rc(k)–Rc_last)-(Sc(k)–Sc_last));0]

The definition of m and k and their relation are the same of the previous section.

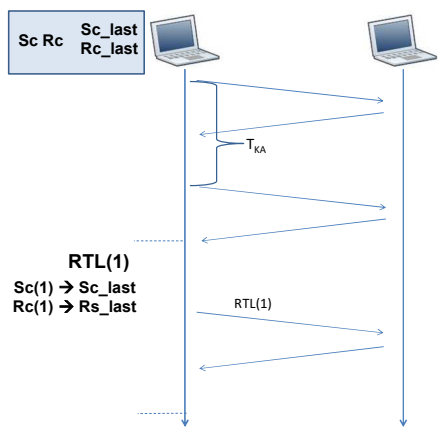

Fig. 20 Round Trip Loss (RTL) evaluation procedure

Note that every time the RTL(m) is evaluated, the counter of received packets Rc_last can be decreased to take into account that during an evaluation interval the number of received packets has been greater than the number of sent packets. In this way the number of received packets in the next evaluation interval will be correspondingly increased. In fact, let us assume that N probe packets are sent over an observation interval $T_L$. Due to the RTT delay a probe reply could be received during the next $T_L$ interval (in this case the algorithm will measure a loss event over the first observation interval). If the RTT remains constant, in the next interval the number of probe requests and probe responses will match and no loss will be detected. If the RTT decreases, one can receive a number of replies larger than N in an observation interval $T_L$. In this case, the excess probe replies received are accounted for in the next interval.

The Round Trip Loss ratio that is evaluated on each interval can be accumulated using an EWMA, as described for the OWL and the RTT. In this case, only another state variable is added per each tunnel.

Overall, for the simplified RTL evaluation, the state variables that need to be maintained in the client-end per each tunnel to be monitored are Rc, Sc, Rc_last, Sc_last, RTL, RTL-EWMA. On the server-end, only the RTL-EWMA state variable needs to be maintained.

## 14 APPENDIX IV: ADDITIONAL RESULTS ON ESTIMATION ACCURACY AND TIMELINESS

In this Appendix we study the relationship between the estimation accuracy and timeliness. Assuming for simplicity that the RTT change in a step-wise manner, we can define the responsiveness of the estimator as the time the EWMA estimate is within a interval of the new value RTT value. More precisely, if we denote $RTT_{old}$ and $RTT_{new}$ the previous and the current value of the RTT, we define the timeliness of the estimate, $T_{tml}$ as the time required for the EWMA estimate to fall within a small interval of new value $RTT_{new}$. More precisely, in order to have a definition which does not depends on the actual value of RTT, we define $T_{tml}$ as the time required for the EWMA to fall within an interval smaller than 10% of the RTT variation, that is, the time required for |EWMA-$RTT_{new}$| ≤ 0.1*| $RTT_{old}$-$RTT_{new}$ | to hold true. If we consider the example in Fig. 21, where the RTT has a sharp rise from 100ms to 200ms, the timeliness of the estimate is the time required for the EWMA to reach the value of 190ms, (10% of the 100ms RTT variation).

We can readily compute $T_{tml}$ from the time constant τ. From the definition, we have that

$$e^{-T_{tml}/\tau} = 0.1$$

from which we obtain

$$T_{tml} = \tau \log 10 \approx 2.3\,\tau$$

As expected the timeliness is directly proportional to time constant τ[2] (and does not depend on the Keep Alive period). We expect that, for timely detection, performance requirement to be expressed in terms $T_{tml}$ from which we derive the time

---

[2] Indeed, while τ is the time for the exponential estimate to settle to 1/e, Ttml is the time for the estimate to settle to 1/10.

constant τ to be used in the EWMA estimate computation.

We repeated the measurements reported in Fig. 16 with a different Keep Alive period. In Fig. 21 $T_{KA}$ = 2 s and the four values of the time constant are: 1.24, 2.18, 3.92, 8.96 s (we used $T_{KA}$ as reference time interval T and the same four values for α).

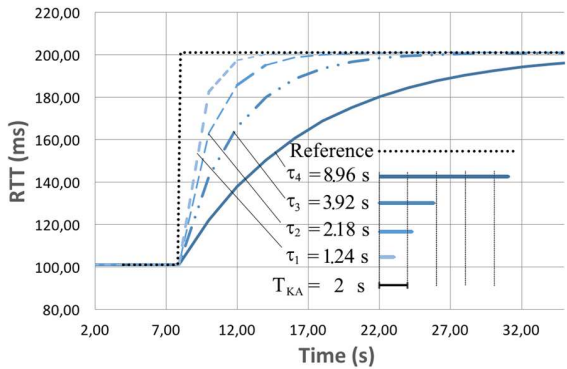

Fig. 21 - RTT step variation ($T_{KA}$ = 2s), different time constants

We repeated the measurements reported in Fig. 17 with a different Keep Alive period. In Fig. 22, $T_{KA}$ = 200 ms, the EWMA is plotted for time constants τ of 124 and 896 ms. In this case, also with the highest considered time constant of 896 ms the EWMA follows quite reasonably the Reference RTT, with a delay in the order of 2-3 s.

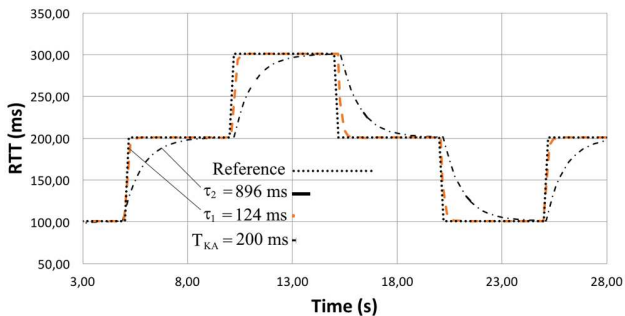

Fig. 22 – RTT 3-levels ($T_{KA}$ = 200 ms) with different time constants

We repeated the measurements reported in Fig. 18 with a different Keep Alive period.

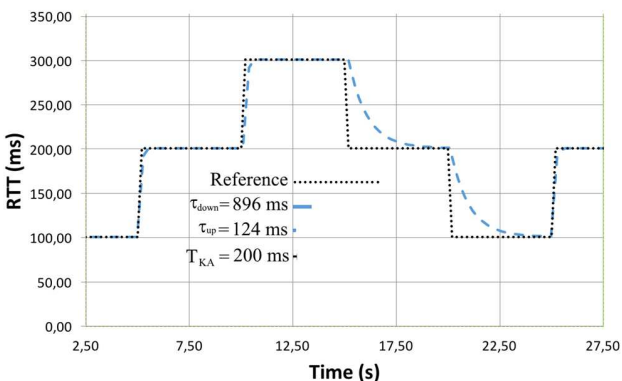

Fig. 23 - Asymmetric RTT approach, $T_{KA}$ = 200 ms

### 14.1 OWL evaluation

Considering the evaluation of loss ratio (OWL), we proceeded in a similar way as in section 6 to verify the functionality and the performance of the implemented solution. We imposed deterministic (and piecewise constant) values for the loss ratios to packets crossing the testbed router and we measured the output of the loss estimation modules. In Fig. 24 the piecewise constant loss ratio was generated according to this periodic profile: 0.1% for 15 s, 10% for 15 s, 20% for 15 s, 10% for 15 s and then start again. According to the definitions given in section 4.2 the loss evaluation interval $T_L$ is equal to $N*T_{KA}$. In these experiment we always set N=10. Hence, for $T_{KA}$ = 200 ms, $T_L$ = 2 s. Fig. 24 show the measured loss ratio (EWMA) compared with the generated loss ratio ("Reference"), for a small time constant τ of 1.24 s, which let the EWMA estimate follow the measured loss ratios with negligible delay.

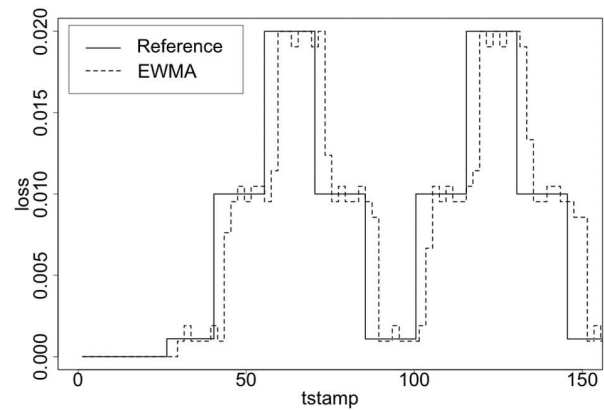

Fig. 24 - OWL 3-levels variation, $T_{KA}$ = 200 ms, N=10, alpha = 0.8